\documentclass[%
 aip,
 amsmath,amssymb,
 reprint,%
]{revtex4-1}

\usepackage{graphicx}
\usepackage{dcolumn}
\usepackage{bm}

\usepackage[utf8]{inputenc}
\usepackage[T1]{fontenc}
\usepackage{mathptmx}
\usepackage{etoolbox}

\usepackage{siunitx}
\usepackage{lipsum}
\DeclareSIUnit\torr{Torr}
\DeclareSIUnit\oersted{Oe}

\makeatletter
\def\@email#1#2{%
 \endgroup
 \patchcmd{\titleblock@produce}
  {\frontmatter@RRAPformat}
  {\frontmatter@RRAPformat{\produce@RRAP{*#1\href{mailto:#2}{#2}}}\frontmatter@RRAPformat}
  {}{}
}%
\makeatother
\begin{document}

\preprint{AIP/123-QED}

\title[Interference patterns of propagating spin wave in spin Hall oscillator arrays]{Interference patterns of propagating spin wave in spin Hall oscillator arrays}

\author{Mohammad Haidar}
 \email{mh280@aub.edu.lb}
 \affiliation{Department of Physics American University of Beirut P.O. Box 11-0236, Riad El-Solh,, Beirut, 1107-2020, Lebanon}

\date{\today}

\begin{abstract}
In this study, we discuss the observation of spin wave interference generated by magnetic oscillators. We employ micromagnetic simulations for two coherent spin Hall nanowire oscillators positioned nearby, horizontally or vertically. The two nanowires produce circular waves with short wavelengths on the order of 100 nm, which interfere with each other. In the horizontal configuration, the spin waves exhibit constructive and destructive fringes, indicating amplification or cancellation of the amplitudes, respectively. The synchronization of spin waves in the current geometry of the two nanowires is facilitated by the combination of
dipolar field and propagating spin waves. Additionally, the vertical alignment results in standing spin waves characterized by multiple antinodes and nodes. These observations are interpreted using a wave model that incorporates the superposition principle for each case.

\end{abstract}

\maketitle


\section{Introduction}
Spin waves, which are low-energy excitations in ferromagnets, have garnered significant attention as potential data carriers in unconventional computation and communication paradigms based on spintronics and magnonic devices \cite{Chumak2015,Mahmoud2020,Papp2021}. Their unique characteristics, including a wide frequency range from gigahertz to terahertz, adjustable wavelength from nano to millimeter scales, and the reconfigurability feature \cite{Haldar2016, Wagner2016, Gladii2016, Albisetti2018} offer potential advantages for magnonic data processing devices \cite{Chumak2014, Khitun2011, Cornelissen2015, Dieny2020}. These devices include RF components such as filters, circulators, phase shifters, multiplexers, and interferometers \cite{Kostylev2005,Schneider2008,Harris2012, Krawczyk_2014, Wang2021}. Most of the developed spin-wave Boolean logic gates operate by utilizing the spin-wave phase \cite{Klingler2014,Liu2011SW,Goto2019, Ustinov2021}. An alternative approach involves encoding data into spin-wave amplitude \cite{Balynsky2017}, utilizing constructive and destructive interference patterns. In this scheme, logic "1" or "0" corresponds to maximum and minimum amplitudes, respectively. To compete with current CMOS technology, spin-wave majority gates must be scaled down to critical dimensions on the order of 100 nm \cite{Talmelli2020}. 

Spin torque oscillators offer a promising alternative pathway for scaling down magnonic devices. Spin torque oscillators have proven to be highly efficient as the smallest microwave emitters, exhibiting rich dynamics of both localized and propagating excitations \cite{DemidovNCSHNO,Ranjbar2014,Haidar2019}. It is widely established that propagating waves are induced in perpendicular magnetic anisotropy (PMA) materials \cite{Houshang2018,Divinskiy2017,Fulara2019, Succar2023}, with frequencies that can be tuned via dc current and magnetic fields. Furthermore, the wavelength of the propagation can be reduced to 100 nm using a straightforward lithography process that is compatible with CMOS technology \cite{Yu2016,Liu2018,Grob2020,HAIDAR2023}. Of notable significance is the ability of propagating spin waves to induce mutual synchronization over micron-scale distances, which holds considerable importance for emerging spin wave computing platforms \cite{Awad2016,Lebrun2017, Litvinenko2023}. Yet, the study of the interference pattern between spin torque oscillators has not been addressed so far.

In this study, we investigate the interference pattern of propagating spin waves induced by spin torque oscillators through micromagnetic simulations. We utilize two coherent spin Hall nanowire oscillators positioned closely, either horizontally or vertically. In the horizontal arrangement, both constructive and destructive spin wave fringes are observed, indicating the amplification or cancellation of interfering waves and resulting in maximum or minimum amplitudes. This setup mimics Young's double-slit experiment. Additionally, in the vertical alignment, the counter-propagation of the two waves results in the generation of standing spin waves characterized by multiple antinodes and nodes. 
These findings are interpreted using a wave model, applying the superposition principle to each case.

\begin{figure*}[t]
\begin{center}
\includegraphics[width=0.9\textwidth]{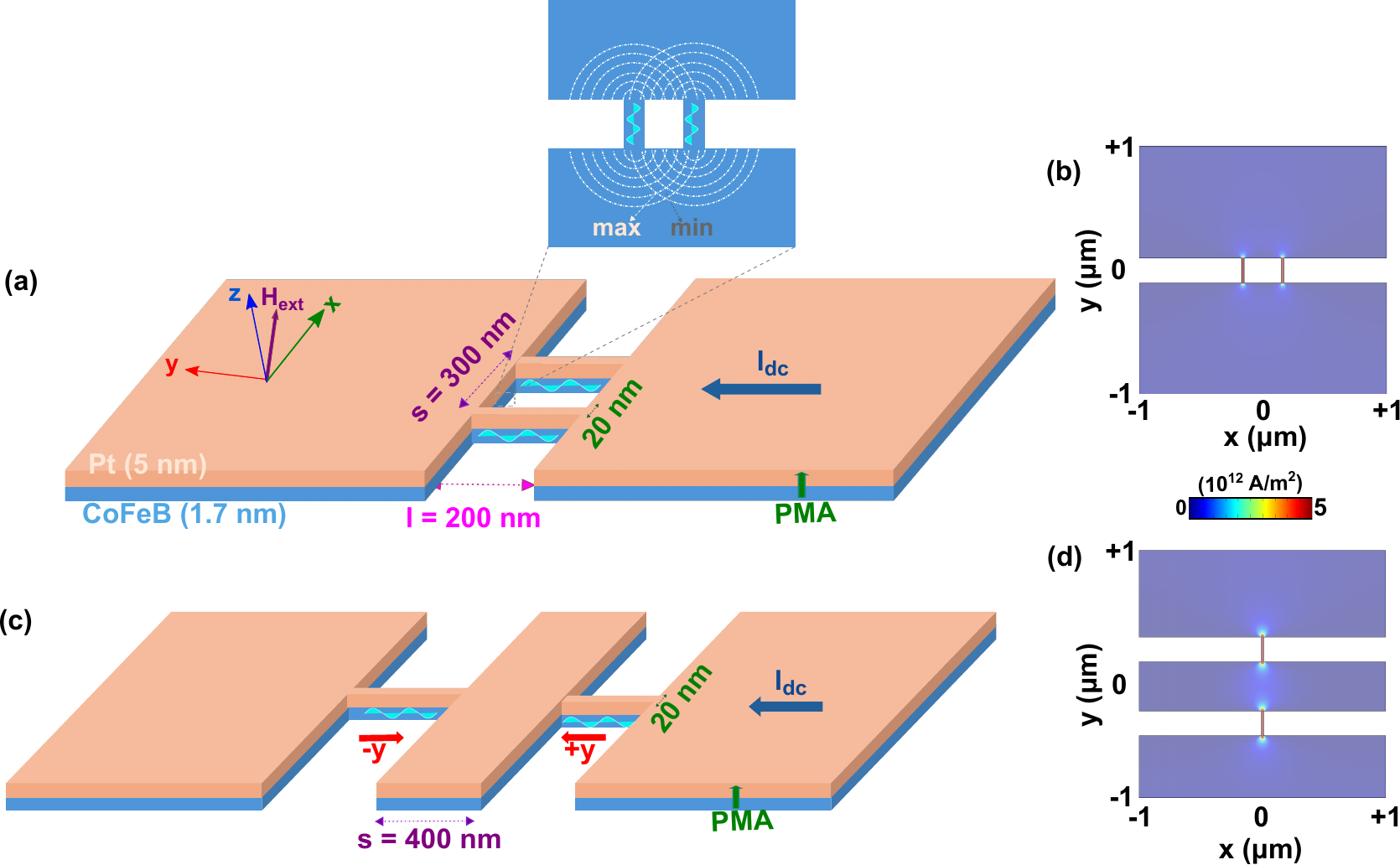}
\caption{(Color online) Schematic representation of the nanowire-based spin-Hall oscillator device made of CoFeB (1.7 nm)/Pt (5 nm) bilayer with two adjacent nanowires of width \textit{w} = \SI{20}{nm}, length \text{l} = \SI{200}{nm} and separation \textit{s} aligned (a) horizontally and (c) vertically. This shows the direction of the applied external magnetic field and the dc current. The inset of (a) shows the interference pattern of two circular waves illustrating maximum and minimum amplitudes. (b, d) The charge current density is calculated at \textit{I} = \SI{2}{mA} using COMSOL software.}.
\end{center}
\end{figure*}

\begin{figure*}[t]
\begin{center}
\includegraphics[width=0.8\textwidth]{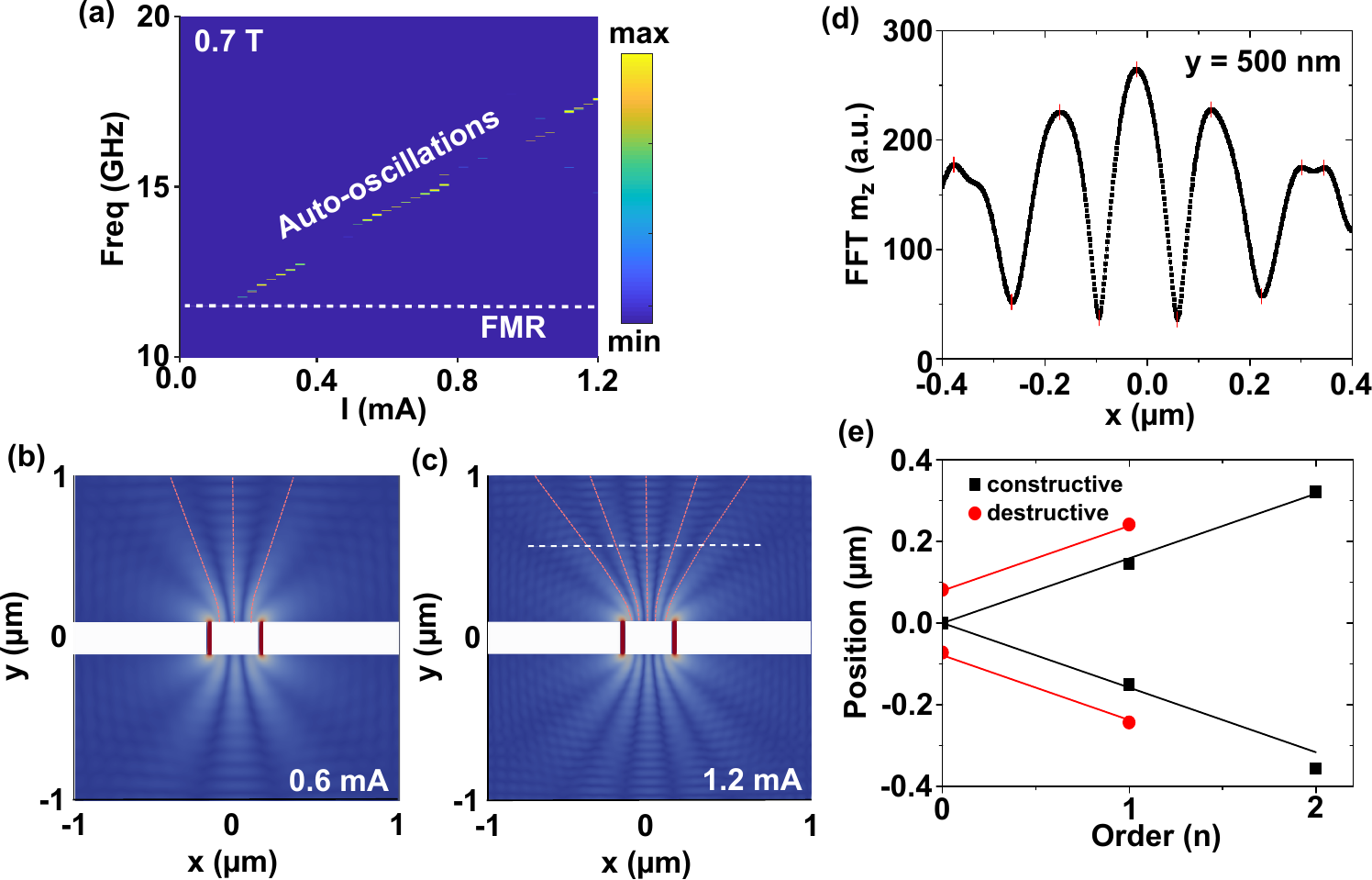}
\caption{(Color online) (a) Power spectra color plot of spin-wave modes in CoFeB/Pt nanowires as a function of dc current. The micromagnetic simulations were conducted under an applied magnetic field of \SI{0.7}{T} at an out-of-plane angle of \ang{75} with a perpendicular magnetic anisotropy  K$_{u} =$ \SI{0.05}{MJ/m^3}. (b, c) Fast Fourier Transform (FFT) of the spatial mode profile of the z-component of the magnetization (\text{m}$_{z}$) for the horizontal alignment of the nanowires at \textit{I} = \SI{0.6}{mA} and \SI{1.2}{mA}. The red dashed lines indicate the constructive fringes. The white dashed line represents the screen positioned at \textit{y} = \SI{500}{nm}. (d) Variation of the FFT of $m_z$ measured at \textit{y} = \SI{500}{nm} along the x-direction for \SI{1.2}{mA}. (e) Loci of the constructive (black symbols) and destructive (red symbols) interference patterns versus the order number. The lines represent the results of fits.}
\end{center}
\end{figure*}

\section{Micromagnetic simulations}
To initiate micromagnetic simulations, we model a layer stack comprising \SI{1.7}{nm} CoFeB and \SI{5}{nm} Pt layers housing rectangular-shaped nanowires with dimensions: length (\textit{l}) = \SI{200}{nm} and width (\textit{w}) = \SI{20}{nm}. We consider two adjacent nanowires arranged horizontally or vertically. For horizontal alignment, the separation distance (\textit{s}) is \SI{300}{nm}, while for vertical alignment, it is \SI{400}{nm} as shown in Fig. 1(a,c). An in-plane current, directed along the y-axis, is applied where a high current density is concentrated within the nanowire region and rapidly diminishes through the electrodes. Using COMSOL software, we simulate the electrical current density and the corresponding Oersted field within the devices with a reference electrical current of \SI{2}{mA}. Fig. 1(b,d) illustrates a plot of the electrical current density for the two devices. Micromagnetic simulations are done using the mumax$^{3}$ solver while integrating input from the COMSOL simulation. In these simulations, we adopt a rectangular mesh with dimensions of \SI{2000}{}$\times$\SI{2000}{}$\times$\SI{1.7}{nm^3} and a cell size of \SI{3.9}{}$\times$\SI{3.9}{}$\times$\SI{1.7}{nm^3}. We convert the electrical current density (\text{J}${e}$) to the spin current density (\text{J}${s}$) via the relation $J_\text{s} = \theta_\text{SH} J_\text{e}$, where $\theta_\text{SH}$ represents the spin Hall angle of Pt and is equal to \SI{0.1}{}. Furthermore, in the simulation, we assume that the injected spin current predominantly induces a damping-like torque of the Slonczewski form \cite{Dvornik2018}.

For micromagnetic simulations, the CoFeB/Pt bilayers are characterized by a saturation magnetization $\mu_{0} M_\text{s}$ of \SI{0.9}{T}, a Gilbert damping coefficient $\alpha$ of \SI{0.02}{}, a gyromagnetic ratio $\gamma/2\pi$ of \SI{30}{GHz/T}, and an exchange stiffness \text{A} of \SI{15}{pJ/m}, consistent with experimental findings \cite{Zahedinejad2018apl}. The magnetization dynamics are simulated by integrating the Landau-Lifshits-Gilbert-Slonczewski (LLG-S) equation over \SI{150}{ns}. Excited mode frequencies and their spatial profiles are determined by performing a Fast Fourier Transform (FFT) of the time-domain data representing magnetization evolution.

\section{Results and Discussion}

\begin{figure*}[t]
\begin{center}
\includegraphics[width=0.8\textwidth]{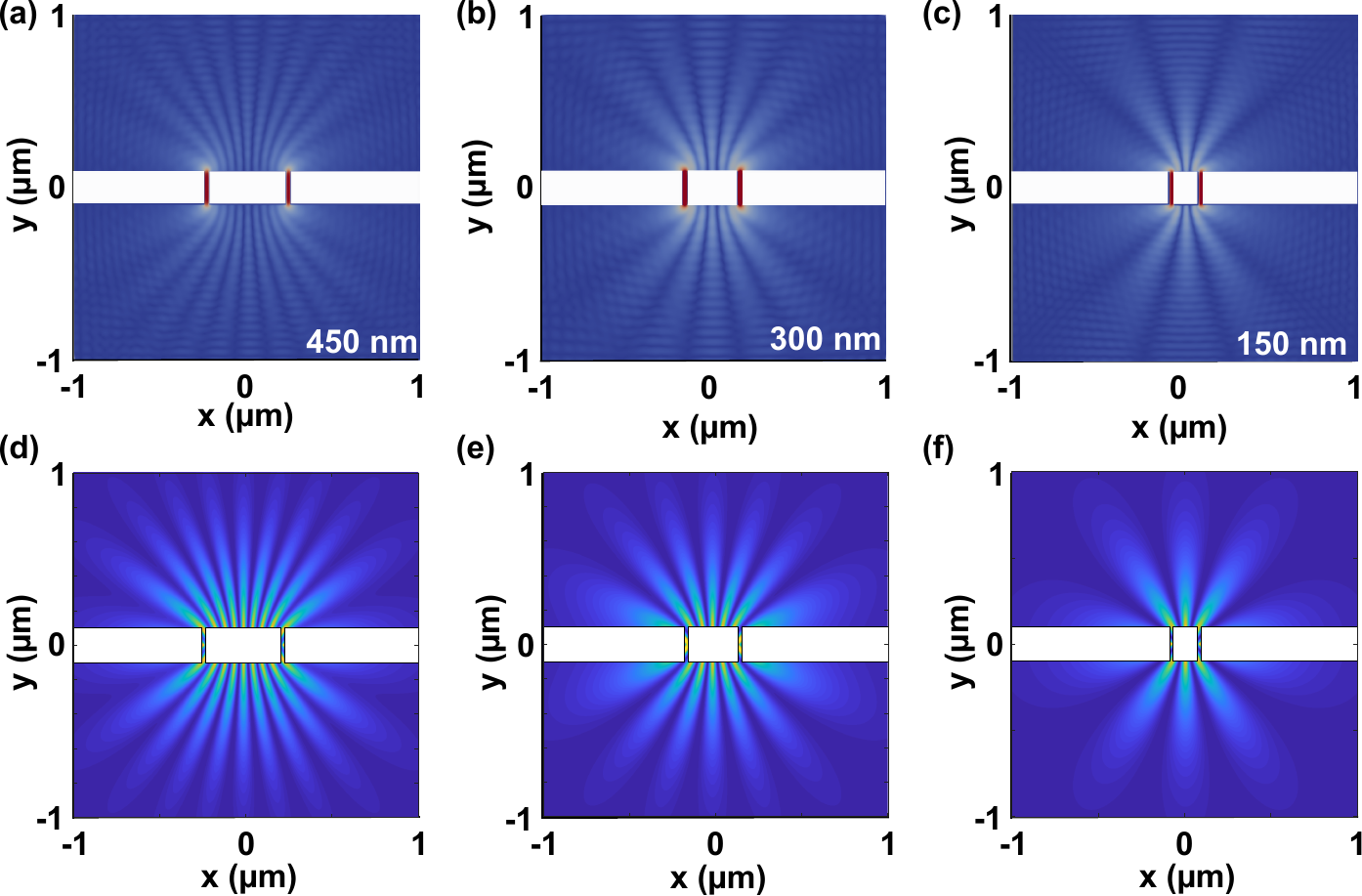}
\caption{Fast Fourier Transform (FFT) of the spatial mode profile of the z-component of the magnetization ($m_z$ ) for varied separation distances (\textit{s}) between the nanowires: (a) \SI{450}{nm}, (b) \SI{300}{n m}, and (c) \SI{150}{nm} simulated at \textit{I} = \SI{1.2}{mA}.  (d-e) Two dimension intensity maps calculated from the superposition between the two oscillating magnetization ($m_1+ m_{2}$) taking into account the attenuation length (\text{L}$_{att}$) =  \SI{500}{nm} for different separation (d) \SI{450}{nm}, (e) \SI{300}{nm}, and (f) \SI{150}{nm}.}
\end{center}
\end{figure*}

The nanowire devices are subjected to an out-of-plane magnetic field of 0.7 T at an angle of $75^{\circ}$. The CoFeB/Pt nanowires are considered to have a perpendicular magnetic anisotropy (PMA) of magnitude $K_{u}=$ \SI{50}{KJ/m^2}. An in-plane dc current is applied along the devices ranging from 0 mA up to 1.2 mA. When the spin polarized current density surpasses the damping torque an emission of auto-oscillations is observed. Figure 2(a) shows the power spectral density of the device as a function of the applied dc current, indicating the emission of auto-oscillations at a threshold current of \SI{0.25}{mA}, where the two identical nanowires oscillate coherently at the same frequency.
Moreover, Fig. 2(a) shows the frequency of the auto-oscillations is higher than the FMR frequency (dashed white line) which indicates an emission of propagating spin wave in the vicinity of the nanowires \cite{Divinskiy2017,Fulara2019, Succar2023}. 
Fig. 2(b,c) shows the FFT of the spatial profile of the magnetization dynamics along the z-direction ($m_{z}$) calculated at 0.6 and 1.2 mA. The two maps reveal the emission of two coherent propagating waves originating from nanowire 1 and nanowire 2 centers and propagating in their vicinity. Interestingly, at a distance r away the nanowires, the two propagating waves interfere either constructively (highlighted by red dashed lines) or destructively depending on their path difference. During constructive interference, the amplitudes of the spin waves increase as the two waves are added when they are in phase. Conversely, during destructive interference, the amplitudes cancel out when the waves are out of phase. These interference patterns exhibit similarities to those observed in Young's double-slit experiment and other investigations of spin-wave interference using coplanar waveguides \cite{Choi2006, Mansfeld2012, Korner2017,Temdie2024}. It is worth noting that more constructive/ destructive fringes are observed at \SI{1.2}{mA} this is due to a shorter wavelengths at higher currents \cite{HAIDAR2023}. Then, we evaluate the location of the constructive and destructive fringes along the x-axis by placing a screen at y = 500 nm (white dashed line). Fig. 2(d) reveals an oscillatory pattern alternating between constructive and destructive fringes for \SI{1.2}{mA}.  
Two observations are noteworthy: (i) the amplitude of the constructive interference fringes slightly diminishes with distance, which is likely due to a finite attenuation length of the spin waves. (ii) The amplitude of the destructive interference approaches zero, likely attributed to a phase difference between the two oscillators. 
The points of maximum constructive interference occur where the crests of the two waves coincide, where the path difference satisfies the condition \cite{Hecht2002}
\begin{equation}
    \Delta r = r_{1}-r_{2}= n\lambda.
\end{equation} 

where $n$ is an integer and $\lambda$ is the wavelength of the propagating spin wave. Conversely, points of destructive interference occur where the crests of one wave align with the troughs of another, resulting in a path difference of 
\begin{equation}
    \Delta r = r_{1}-r_{2} = (2n+1)\lambda/2
\end{equation}
with $n$ being an integer (0, 1, 2, 3,...). 
Fig. 2(e) shows the position of constructive (black squares) and destructive (red circles) fringes as a function of the order $n$ along with fitting the path difference for both fringes. Fitting the data simultaneously we extract the wavelength $\lambda$ of $150$ and $95$ nm for the 0.6 and 1.2 mA respectively similar to the value reported in \cite{HAIDAR2023}. Notably, a similar interference pattern is observed when modifying the magnitude of the PMA, primarily affecting the magnitude of the wavelength.

Furthermore, we examine the effect of the separation distance between the two oscillators on the interference pattern. We perform the simulations using an out-of-plane (OOP) magnetic field of 0.7 T and an in-plane current of 1.2 mA. Fig. 3 shows the FFT of the spatial profile for three separation (\textit{s}) of 150, 300, and 450 nm as shown in Fig. 3 (a-c). Similar constructive and destructive interference fringes are observed for all devices. However, we notice the following remarks: (i) performing similar analysis as that mentioned above we find the wavelength for devices with \SI{450}{nm} and \SI{300}{nm} is \SI{95}{nm}. However, we estimate a shorter wavelength of \SI{85}{nm} for a separation distance of \SI{150}{nm} which could be attributed to a stronger coupling between the adjacent oscillators.  
(ii) The number of fringes increases with the increase of the separation \textit{s}. The number of fringes is determined by the condition: $n < \frac{s}{\lambda}$, where $n$ is an integer. If $s$ is less than or equal to $\lambda$, only the central bright spot will be observed. 
In the present case, we observe one, three, and five bright fringes for the 150, 300, and 450 respectively. (iii) For these devices, the frequency of the two oscillators is perfectly synchronized, and they oscillate in phase. However, for larger separations above 600 nm, the frequency of the two oscillators is not synchronized, resulting in no interference pattern being observed. 
The synchronization of spin-torque oscillators is mediated by (i) dipolar coupling, and (iii) propagating spin waves \cite{Slavin2006, slavin2009, Kendziorczyk2014, Kendziorczyk2016}. Dipolar coupling exhibits a rapid decay with the distance. However, propagating spin waves can induce mutual synchronization over longer distances, although their effects are predominantly influenced by the finite attenuation length. Thus, we can infer that the synchronization of spin waves in the current geometry of the two nanowires is facilitated by the combination of dipolar field and propagating spin waves. However, the interference pattern arises from spin wave propagation, adhering to the classical pattern of constructive/destructive interaction between two waves.

\begin{figure*}[t]
\begin{center}
\includegraphics[width=0.6\textwidth]{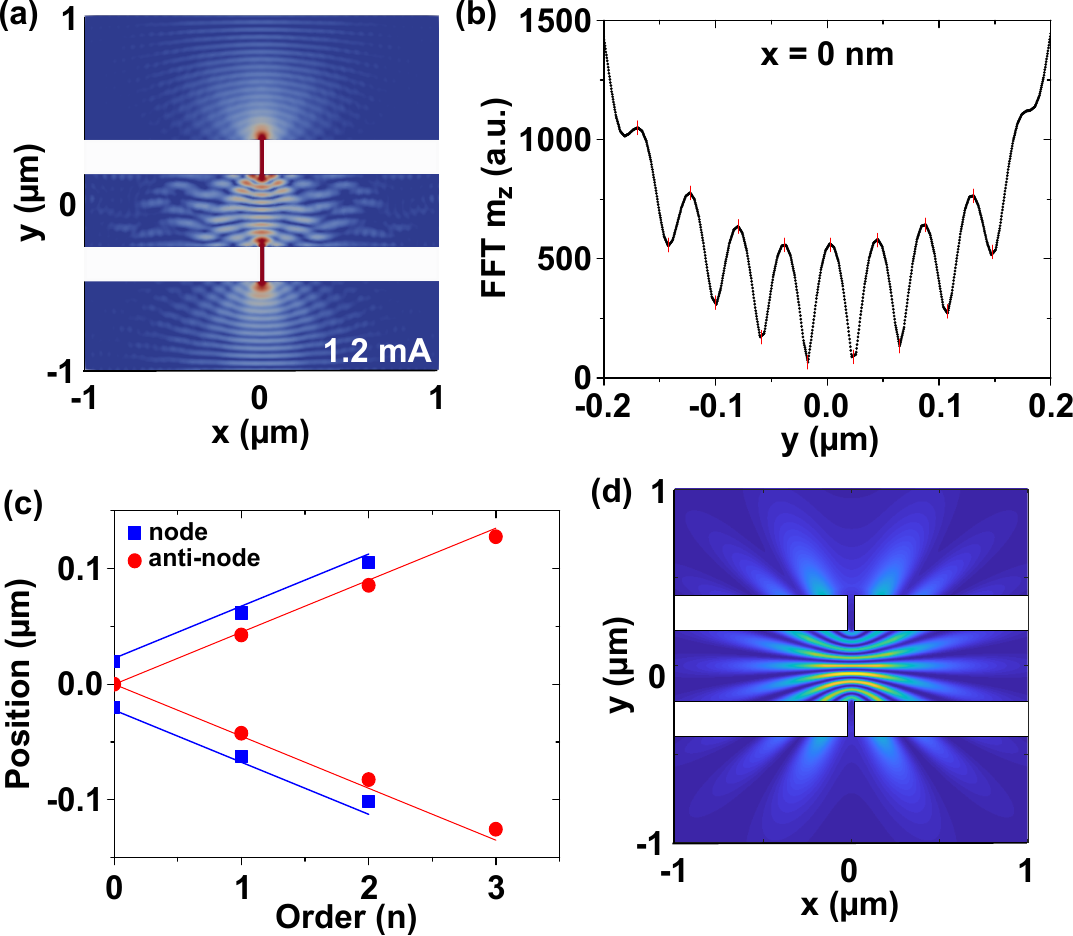}
\caption{(a) Fast Fourier Transform (FFT) of the spatial mode profile of the z-component of the magnetization (\text{m}$_{z}$) for the vertical alignment of two nanowires separated by a distance of \SI{400}{nm}, simulated at \textit{I} = \SI{1.2}{mA}. (b) Variation of the FFT of $m_z$ measured at \textit{x} = \SI{0}{nm} along the y-direction. (c) Loci of the nodes (blue symbols) and anti-nodes (red symbols) versus the mode order. The lines represent the results of fits. (d) Two-dimensional intensity maps is calculated from the superposition of the two oscillating magnetizations ($m_1$ - $m_{2}$).}
\end{center}
\end{figure*}
Therefore, we interpret the micromagnetic simulation using a wave model, expressing the magnetization of each microwave source in terms of a propagating wave as follows \cite{Loayza2018}:

\begin{equation}
m_{1,2}= A e^{ik.r_{1,2}}. e^{-r_{1,2}/L_att}
\end{equation}

Here, A represents the amplitude of the wave, k is the wavevector, and $r_{1,2}$ is the distance to the nanowires calculated as $\sqrt{((x-x_{1,2})^2+y^2)}$ where $x_{1,2}$ are the position of the nanowires. The last term is incorporated to consider the decaying factor of the spin waves with an attenuation length ($L_{att}$) of \SI{500}{nm}. We apply the superposition principle for the two wave sources by taking the sum of $m_1$ + $m_{2}$ and calculate the resultant intensity over $2\times 2$ $\mu m$ map. In the model, we employed values of the wavelength determined from the micromagnetic simulations, and we positioned the screen at \SI{500}{nm}. Figures 3(d-f) show the intensity maps generated using the superposition principle. A perfect correspondence between the superposition model and the micromagnetic simulation is evident: (i) The model accurately reproduces both bright and dark fringes. (ii) It precisely models the exact number of fringes. (iii) The intensity decreases as a function of distance.

Furthermore, we analyze the interference pattern of the two vertically aligned nanowires separated by a distance of 400 nm, as shown in Fig. 4. Fig. 4(a) shows the FFT of the spatial profile of magnetization oscillation along the z-direction ($m_{z}$). In this arrangement, one oscillator generates a propagating wave traveling in the positive y-direction, while the other produces a wave in the negative y-direction. Between the two oscillators, a standing spin wave (SSW) is formed with observable antinodes and nodes. To determine the positions of fringes, we scan the variation of $m_{z}$ along the x = 0 line, shown in Fig. 4(b). Several observations can be made: (i) The maximum amplitude is observed near the two oscillators (-0.2 and 0.2 $\mu m$), and gradually decreases away from the microwave source. (ii) The magnitude of the antinode remains relatively constant around the midpoint between the two oscillators. (iii) The magnitude of the node decreases, nearly reaching zero near the center between the two oscillators, resulting in a perfect node. Fig. 4(c) shows the positions of antinodes (closed circles) and nodes (closed squares) as a function of the mode order. By considering the boundary condition to be maximum near the nanowires' edges, we determine the positions of antinodes and nodes to correspond, respectively, \cite{Hecht2002} 
\begin{align} 
y= n\lambda/2 \\
y= (2n+1)\lambda/4
\end{align} 
where $n$ is an integer. The solid lines in Fig. 4(c) are fitted to Eqs. (4) and (5) with a wavelength ($\lambda$) of 90 nm, demonstrating an excellent match between the simulated results and the calculated positions. Using the wave model, discussed above, we reproduce the interference pattern for this configuration by subtracting the waves generated by the two sources ($m_1$ - $m_{2}$), utilizing a wavelength of 90 nm. 
The $2 \times 2$ $\mu m$ map intensity is shown in Fig. 4(d) where one can observe that the number of fringes and the positions of the constructive and destructive fringes are accurately reproduced. This confirms the formation of standing waves in the vicinity of the nanowires. 

\section{\label{sec:level1}Conclusion}

In conclusion, the interference behavior of propagating spin waves is explored through micromagnetic simulations. Employing a simplified device comprising an array of spin Hall oscillators, we demonstrated the excitation and the interference of propagating spin waves driven by spin torque. The alignment of the two oscillators determines whether constructive, destructive, or standing waves are generated. These findings hold significant implications for the development of spin wave-based logic devices, where both amplitude and phase play crucial roles.

\textbf{Acknowledgment}
\\ 
This work is supported by the American University of Beirut Research Board (URB).

\textbf{DATA AVAILABILITY}

The data that support the findings of this study are available from the corresponding author upon reasonable request.

\end{document}